\documentclass[twocolumn]{pasj01}
\Received{$\langle$reception date$\rangle$}
\Accepted{$\langle$acception date$\rangle$}
\Published{$\langle$publication date$\rangle$}

\begin{document}

\title{
Pulsed fraction of super-critical column accretion flows onto neutron stars:
modeling of ultraluminous X-ray pulsars
}
\author{Akihiro \textsc{Inoue},\altaffilmark{1,}$^{*}$
Ken \textsc{Ohsuga},\altaffilmark{1}
Tomohisa \textsc{Kawashima},\altaffilmark{2}}%
\altaffiltext{1}{Center for Computational Sciences, University of Tsukuba, Ten-nodai, 1-1-1 Tsukuba,
Ibaraki 305-8577, Japan}
 \altaffiltext{2}{Center for Computational Astrophysics, National Astronomical Observatory of Japan, 2-21-1 Osawa, Mitaka, Tokyo 181-8588, Japan}
\email{akihiro@ccs.tsukuba.ac.jp}


\maketitle

\begin{abstract}
We calculate the pulsed fraction (PF)
of the super-critical column accretion flows onto magnetized neutron stars (NSs),
of which the magnetic axis is misaligned with the rotation axis,
based on the simulation results by Kawashima et al.(2016,~PASJ,~68,~83).
Here, we solve the geodesic equation for light in the Schwarzschild spacetime
 in order to take into account the light bending effect.
 The gravitational redshift and the relativistic doppler effect from
 gas motions of the accretion columns are also incorporated.
The pulsed emission appears since the observed luminosity, 
which exceeds the Eddington luminosity for the stellar-mass black holes,
periodically changes via precession of the column caused by the rotation of the NS.
The PF tends to increase as $\theta_{\rm obs}$ approaching to $\theta_{\rm B}$,
where $\theta_{\rm obs}$ and $\theta_{\rm B}$ are
the observer's viewing angle and the polar angle of the magnetic axis
measured from the rotation axis.
The maximum PF is around $50$ \%.
Also, we find that the PF becomes less than 5 \%
for $\theta_{\rm obs} \lesssim 5^\circ$ or for $\theta_{\rm B}  \lesssim 5^\circ$.
Our results are consistent with observations of
ultraluminous X-ray pulsars (ULXPs) with few exceptions,
since the ULXPs mostly exhibit the PF of $\lesssim 50$ \%.
Our present study supports
the hypothesis that the ULXPs are powered by
the super-critical column accretion onto NSs.
\end{abstract}

\section{Introduction}
The ultraluminous X-ray sources (ULXs) are off-nuclear, compact, X-ray sources
of which the X-ray luminosity exceeds the Eddington luminosity
for the stellar-mass black holes, $\sim 10^{39}~{\rm erg~s^{-1}}$.
The central engine and energy production mechanism of the
ULXs are hottly debated issues.
It has been thought to be either sub-critical accretion disk
around an intermediate-mass black hole \citep{Colbert, Makishima, Kobayashi}
or super-critical accretion flow onto a stellar-mass black hole
\citep{King01, Watarai, Vierdayanti, Poutanen}.
Super-critical accretion scenario has been supported
by the radiation hydrodynamics/radiation-magnetohydrodynamics simulations
which revealed that the disk luminosity
becomes larger than the Eddington luminosity
in super-critical accretion regime
(\cite{Ohsuga2005}, \yearcite{Ohsuga2009}; \cite{Ohsuga2011}; \cite{Vinokurov}; \cite{Jiang}; \cite{Sadowski}; \cite{Takahashi16}).
However, the central objects of some ULXs turned out to be neutron stars (NSs)
since recent X-ray observations detected the pulsed emission.
These ULXs are called ULX Pulsar (ULXP).
It has been pointed out that
a large fraction of ULXs might contain NSs
(\cite{Kluzniak}; \cite{King17}; \cite{Wiktorowicz}; \cite{Pintore}).
Because the mass of the NS is less than a few $M_\odot$,
the matter should accrete at the super-critical rate.
The pulsed fraction (PF), which is defined as
$(L_\mathrm{max}-L_\mathrm{min})/(L_\mathrm{max}+L_\mathrm{min})$
with $L_\mathrm{max}$ and $L_\mathrm{min}$ being the
the maximum and minimum X-ray luminosities on the pulse phase,
has been reported to increase with an increase of photon energy
and to be a few percent to several tens of percent
(see e.g., \cite{Bachetti}; \cite{Furst}; \cite{Israela}, \yearcite{Israelb}).

One of the plausible models to explain both super-Eddington luminosity
and pulsed emission is super-critical column accretion flows
onto a NS, which was initially proposed by \citet{Basko76}.
The accretion disks around the magnetized NSs
are thought to truncate without connecting to the NS surface.
This is the case that the accreting gas
moves along the magnetic field lines
and falls around the magnetic poles of the NSs.
In this way, the accretion column forms around the magnetic poles.
The inflowing gas within the column
is heated up at the vicinity of the NS surface via the shock,
and a large number of photons are emitted
from the side wall of the column.
If the magnetic axis is misaligned
with the rotation axis of the NSs,
the observed luminosity periodically changes
by precession of the column.

Recently, the super-critical accretion column model
has been supported by \citet{Kawashima16} (hereafter K16).
They revealed that the luminosity of the side wall
of the column exceeds the Eddington luminosity.
In their simulations,
the shock appears at the height of about 3 km above the NS surface.
The energy of the infalling matter
is converted into radiation below the shock surface,
and such converted energy is radiated away from the side wall
of the column.
Thus, the column luminosity of the region
very close to the NS surface ($\lesssim 3$ km)
exceeds the Eddington luminosity.
However,
it hasn't been investigated yet
whether the super-critical column accretion
can explain the PF observed in ULXPs or not.

In this paper,
we calculate the observed luminosity of the super-critical accretion column
and the pulse shape (time variation of the observed luminosity caused
by the precession of the columns via the rotation of the NSs) based on
the simulation results of K16.
Here, the light bending effect, the doppler effect,
and the gravitational redshift are taken into consideration.
We investigate the PF
for a wide variety of the observer's viewing angle
and the offset angle between the magnetic axis and the rotation axis.
In section 2, our model and numerical method are described.
We present our results in section \ref{sec:Result}, and section \ref{sec:discussion} is devoted to conclusions and discussion.

\section{Model and Method}
In the present study,
we calculate the pulse shape of the observed luminosity of the super-critical accretion column
and investigate the PF.
The light bending effect is considered by solving the geodesic equation for light.
Since the gas falls onto the NS with the velocity of $\gtsim$ a few 10 \% of the light velocity,
we take the doppler effect into consideration.
The gravitational redshift is also included in our calculations.
The details of our model and calculation method are described below.

\subsection{Model of super-critical accretion column}
\label{sec:model}
\begin{figure}
  \centering
  \includegraphics[width=6cm]{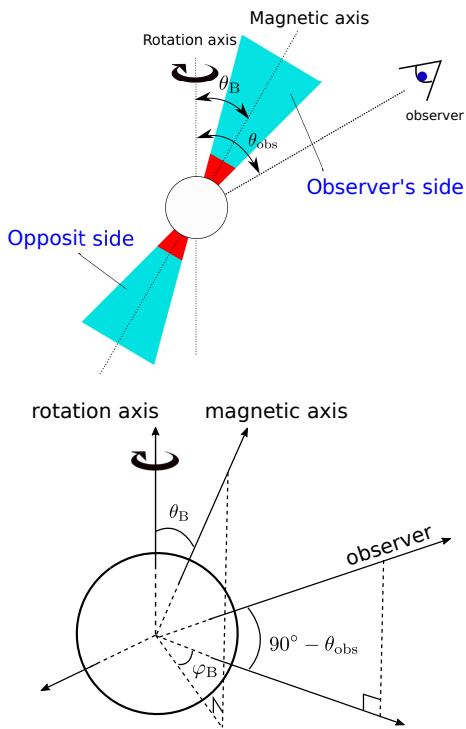}
  \caption{Top: Schematic view of the super-critical accretion column
    onto the neutron star. Bottom: The definition of
    polar and azimuthal angles of the magnetic axis of the NS
    ($\theta_{\rm B}$ and $\varphi_{\rm B}$), and observer's viewing angle ($\theta_{\rm obs}$).}
  \label{fig:Schematic}
\end{figure}

The fiducial model of the super-critical accretion column employed in the present study
is based on the numerical results by K16,
in which the radiation hydrodynamics simulations were performed.
The schematic view of the model is shown in figure \ref{fig:Schematic}.
The mass of the NS is $1.4 M_\odot$ and its radius is 10 km.
The accretion columns appear along the magnetic axis of the NS.
According to K16, the half opening angle of the column, $\Theta_{\rm op}$, is set to be 30 degrees.
The accretion column is assumed to be steady
and axisymmetric with respect to the magnetic axis.
In K16, the column is very optically thick
so that we suppose that the side wall of the column is the photosphere,
emitting the radiation of which the intensity is $\propto T^4$ in the comoving frame,
where $T$ is radiation temperature.
The relativistic correction for the emission
is taken into consideration with using the infall velocity
of the side wall of the column.

\begin{figure}
  \centering
  \includegraphics[width=8cm]{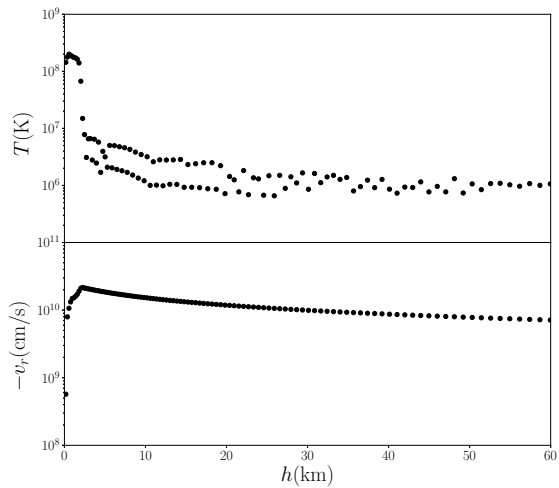}
  \caption{Radiation temperature (top) and infall velocity (bottom)
    at the side wall of the accretion column
    obtained by the radiation hydrodynamics simulations
    of K16.
    Here, $h$ is the distance from the neutron star surface.}
   \label{fig:Kawashima}
  \end{figure}

Figure \ref{fig:Kawashima} shows the radiation temperature, $T$, and the infall velocity, $-v_r$,
of the side wall of the accretion column in the quasi-steady state,
which are obtained by K16.
Here, $h$ is the distance from the neutron star surface.
As shown in this figure, the radiation temperature is effectively enhanced, $>10^8$ K,
at the region of $h\lesssim 3~{\rm km}$, while we find
very small radiation temperature, $\sim 10^{6-7}$ K, at $h\gtsim 3~{\rm km}$.
Thus, the region of $h\lesssim 3~{\rm km}$ of the column is main radiation source
in the model (red part of figure \ref{fig:Schematic}).
Here we note that
the numerical fluctuation of radiation temperatures in the region of $h\gtsim 3~{\rm km}$
does not affect our results since the emission is negligibly weak because of the very small $T$.
The high $T$ region is produced by the shock.
Indeed, the infall velocity is suddenly decreasing within $\sim 3~{\rm km}$, i.e., in the post-shock region.
The intrinsic luminosity of the columns, which is evaluated as $L_{\rm int}=4\pi\int_{r_{\rm in}}^{r_{\rm out}}\sigma T^4 r\sin \Theta_{\rm op} dr \sim 6.1 \times 10^{40}~{\rm erg~s}^{-1}$,
where $\sigma$ is the Stefan-Boltzmann constant, $r$ relates to $h$ as $r = 10 {\rm km} + h$, $r_{\rm in}$ is 10 km (the radius of the NS), and $r_{\rm out}$ is 2100 km (the upper boundary of the simulation box of K16). As we have mentioned above, the most of photons are emitted from the very vicinity of the neutron stars, $h \lesssim 3$ km.

We consider that the magnetic axis (axis of the column) is misaligned with the rotation axis of the NS.
Thus, the precession of the column occurs via the rotation of the NS,
which leads to periodically change of the observed luminosity (pulsed emission).
The polar and the azimuthal angles of the magnetic axis are represented by $\theta_{\rm B}$
and $\varphi_{\rm B}$,
where $\theta_{\rm B}$ is measured from the rotation axis of the NS
and $\varphi_{\rm B}$ is measured from the observer's azimuthal angle
(the observer's viewing angle is defined as $\theta_{\rm obs}$).
That is, the magnetic axis is the closest to the observer
when $\varphi_{\rm B}$ is null.
Note that $\varphi_{\rm B}$ corresponds to the phase of the pulsed emission since $\varphi_{\rm B}$
changes with time via the rotation of NS.
Throughout the present study,
the column which is located at the upper (lower) side in figure \ref{fig:Schematic}
is called observer's side column (opposite side column).
Since the observed luminosity of the columns depends on the positional relationship between the observer and the columns, it is functions of  $\theta_{\rm obs},~\theta_{\rm B}$, and $\varphi_{\rm B}$.

In our method,
we suppose that the NS surface does not emit photons,
since the emission of the NS surface is thought to be
negligibly small, compared with the side wall of the columns.
We neglect the reflection of photons at the NS surface
for simplicity.
Hence, if the photons from the accretion columns arrive at the NS surface,
such photons disappear.
Also we assume the gas outside of the NS and the accretion columns is very optically thin
(the opacity is negligibly small).

In addition to the fiducial model described above,
we calculate two models for comparison.
One of which is a thin column model.
In this model,
the half opening angle of the column is 10 degrees.
In the other model, the hot and luminous regions of the side wall extend
twice (the shock surface is located at $h \sim 6$ km).
These comparison models are described in subsection \ref{sec:comparison models} in detail.

\subsection{Numerical Method}
\label{Numerical Method}
In order to evaluate the observed luminosity of the accretion columns with $\theta_{\rm B}$ and $\varphi_{\rm B} $
for the observer with $\theta_{\rm obs}$, we compute the intensity of radiation, incidenting on the observer screen
placed far enough from the neutron star, by using a ray-tracing method (see, e.g., \cite{Luminet}).
The size of the screen is $-21 {\rm km}<y<21 {\rm km}$ in the vertical
direction and $-21 {\rm km}<x<21 {\rm km}$ in the horizontal direction.
Here, the $y$ axis is defined to be identical to the rotation axis of the neutron
star projected onto the screen.
The $x$ axis, which is perpendicular to the $y$ axis, is defined in such a way
that the center of the NS is projected on the origin of the screen $(x=y=0)$.
This screen is divided into $500 \times 500$ pixels ($\Delta x=\Delta y=84$ m),
and trajectories of the light ray are calculated from each pixel towards the NS.
We assume the Schwarzschild metric outside the NS.
This assumption will be appropriate, since the rotation period of
NSs in ULXPs is the order of seconds
(see e.g., \cite{Bachetti}; \cite{Furst}; \cite{Israela}, \yearcite{Israelb}),
i.e., the expected rotation velocity
of the NS surface is very slow ($\sim 10^{-4} c$).

A light ray from a pixel at $x=x'$ and $y=y'$
travels on the plane including three points:
two points on the screen [$(x,y)=(0,0)$ and $(x',y')$] and the center of the NS,
because of the spherical symmetry of the Schwarzschild spacetime.
Thus, the trajectory of light is obtained by solving
the geodesic equation on the plane in the circular coordinates ($r, \xi$),
\begin{eqnarray}
 \left[\frac{d}{d\xi}\left(\frac{1}{r}\right)\right]^2+\frac{1}{r^2}\left(1-\frac{r_\mathrm{s}}{r}\right)=\frac{1}{b^2},
 \label{eq:geodesic}
\end{eqnarray}
with $r_{\rm S}$ being the Schwarzschild radius (see e.g., \cite{Misner}).
Here, r is the distance from the center of the NS.
The angle, $\xi$, is measured from the line
that links the center of the NS ($r=0$) to the origin of the screen, $x=y=0$.
The impact parameter, $b$, is defined as $b^2=x'^2+y'^2$.

When a ray reaches the side wall of the accretion column,
the radiation temperature, $T$, and the infall velocity, $-v_r$,
at that point are obtained by linear interpolation
from the simulation data of K16 (figure \ref{fig:Kawashima}).
Then, the frequency-integrated intensity of the radiation at the point on the observer's screen,
$x=x'$ and $y=y'$, is evaluated as
\begin{eqnarray}
 I_\mathrm{obs}=\frac{B_0}{(1+z)^4}.
 \label{eq:radiative}
\end{eqnarray}
Here, $B_0$ is black body intensity in the comoving frame of the fluid,
$B_0=\sigma T^4/\pi$ with $\sigma$ being the Stefan-Boltzmann constant.
The radshift, $z$, is defined as
\begin{eqnarray}
 1+z=\left(1-\frac{r_\mathrm{s}}{r}\right)^{-1/2}\gamma
 \left(1-\frac{\boldsymbol{v}_r\cdot\boldsymbol{n}}{c}\right),
 \label{eq:redshift}
\end{eqnarray}
where $c$ is the speed of light,
$\gamma=\left(1-v_r^2/c^2 \right)^{-1/2}$ is the Lorentz factor,
$\boldsymbol{v_r}$ is the velocity vector at the side wall of the column, of which the radial component is -$v_r$ and other components are zero,
and $\boldsymbol{n}$ is the directional cosine of the light ray
at the side wall of the accretion column (\cite{BH-disk}).
Here, $\left( 1-r/r_{\rm S}\right)^{-1/2} $ represents the gravitational redshift,
and $\gamma \left(1-\boldsymbol{v}_r \cdot \boldsymbol{n}/{c} \right)$
corresponds to the Doppler effect.
In the present study, we ignore the relativistic effect caused by
the rotation of the NS $(\sim 1~{\rm s})$, since the rotation speed is much slower
than the infall velocity.
In addition, the rotation period of the NS is thought to be
much longer than the light crossing time around the NS,
$\sim 10^{-4}$ s.
Thus, we do not need to consider the movement of the accretion column
when we solve the trajectories of the light rays.

If the geodesic reaches the surface of the NS prior to the accretion column or
the light ray goes away without reaching the accretion column,
then we recognize the radiation intensity on the pixel is null, $I_\mathrm{obs}=0$.
The radiation flux can be obtained from $I_\mathrm{obs}$ for all pixels.
The observed luminosity is evaluated
by assuming that the radiation flux is isotropic.
By repeating the above procedures with varying $\varphi_{\rm B}$,
we obtain the pulse shape and the PF.

\section{Result}
\label{sec:Result}
  \subsection{Intenisty Map}
  \label{sec:Imaging}
  \begin{figure*}
      \centering
      \includegraphics[width=17cm]{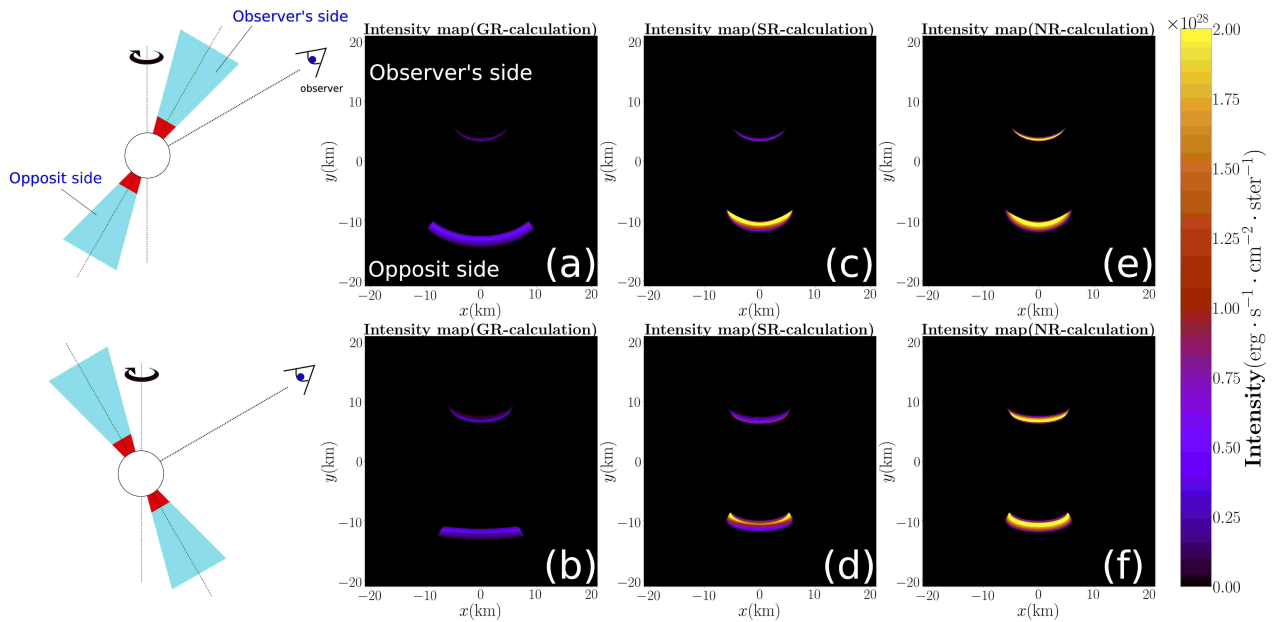}
      \caption{Intensity map of observer's screen for
        $\varphi_{\rm B}=0^\circ$ (upper panels) and $180^\circ$ (lower panels).
        Here, we employ $\theta_\mathrm{B}=10^\circ$ and $\theta_\mathrm{obs}=60^\circ$.
        The neutron star is located at the center of the screen.
        Also, the observer's side (opposite side) column appears
        on the upper (lower) part of the screen.
        The general and special relativistic effects are included
        in the GR-calculation (panels a and b), and only
        the special relativistic effect is considered
        in the SR-calculation (panels c and d).
        In the NR-calculation, the relativistic effects are neglected (panels e and f).}
      \label{fig:Imaging}
  \end{figure*}

The panels (a-f) in figure \ref{fig:Imaging} show the resulting intensity map of the fiducial model
for $\theta_\mathrm{B}=10^\circ$ and $\theta_\mathrm{obs}=60^\circ$.
The azimuthal angle of the magnetic axis is $\varphi_{\rm B}=0^\circ$ in the upper panels
and $\varphi_{\rm B}=180^\circ$ in the lower panels.
The left panels are the schematic picture for $\varphi_{\rm B}=0^\circ$ (upper left) and for $\varphi_{\rm B}=180^\circ$ (lower left).
The accretion column of observer's side is tilted to the observer side at $\varphi_{\rm B}=0^\circ$.
On the other hand, the accretion column tends to be viewed from the side when $\varphi_{\rm B}=180^\circ$.

The panels (a) and (b) are results obtained by the calculation method in subsection \ref{Numerical Method}.
That is, the light bending effect, the gravitational redshift, and the Doppler effect are fully taken into consideration.
We refer to this type of calculations as the General Relativistic calculation (GR-calculation)
to distinguish it from the other calculations.
As can be seen in the panels, two luminous parts (purple) appear.
They correspond to the luminous part of accretion columns near the NS surface $h \lesssim 3 {\rm km}$
(red part in figure \ref{fig:Schematic}).
The panels show that
the intensity of the luminous part of the opposite side is stronger than that of the observer's side.
Such a difference of the brightness is more pronounced
for $\varphi_{\rm B}=0^\circ$ than for $\varphi_{\rm B}=180^\circ$.
Furthermore, in the case of $\varphi_{\rm B}=0^\circ$,
it is also found that the area of the luminous part of opposite side is seen widened in the arc shape.
Due to the enhancement of the intensity and the spread of the luminous area on the screen,
the observed luminosity is larger for $\varphi_{\rm B} = 0^\circ$ than for $\varphi_{\rm B} = 180^\circ$.

The reason why the radiation intensity on the opposite side
looks stronger than that on the observer's side is the Doppler effect.
In order to make the point clear,
we calculate the radiation intensity by neglecting the gravitational redshift
and the light bending effect.
Specifically, we take the limit of $r_{\rm S}/r \rightarrow 0$
in equations (\ref{eq:geodesic}-\ref{eq:redshift}).
We name this type of calculations the Special Relativistic calculation (SR-calculation)
and the resulting intensity map is shown in the panels (c) and (d).
Since neither the general relativistic effect
(light bending and gravitational redshift)
nor the special relativistic effect (Doppler effect) is included in panels (e) and (f),
We call this type of calculations the Non-Relativistic calculation (NR-calculation).

In the panel (c), we find that the radiation intensity of the accretion column
is stronger on the opposite side than on the observer's side.
This is caused by the Doppler effect.
Since the gas in the column accretes onto the NS ($v_r < 0$),
$\boldsymbol{v}_r \cdot \boldsymbol {n}$ becomes negative on the observer's side.
Thus, the radiation intensity, emitted from the observer's side, on the observer's screen is weaker than the
black body intensity in the comoving frame.
In contrast, the column of the opposite side seems to be bright
because of $\boldsymbol{v}_r \cdot \boldsymbol{n} > 0$.
On the other hand, the both sides of the column seem to be bright in the panel (e).
This proves that the Doppler effect is mainly responsible for the
difference of the radiation intensity between the opposite and observer's sides.
In the panel (d), the column tends to be observed from the side
(see lower-left schematic view).
The accretion columns are cone-shaped so that the matter at the side wall
moves away from the observer.
Thus, $\boldsymbol{v}_r \cdot \boldsymbol{n}$
becomes negative in both sides of the column,
and the radiation intensity of the both sides of the accretion column
is weaker in the panel (d) than in the panel (f).
Note that even for the case of $\boldsymbol{v}_r \cdot \boldsymbol{n}=0$,
the radiation intensity is reduced via the Doppler effect since the redshift is
proportional to $\gamma$ (see equation (\ref{eq:redshift})).

In the GR-calculation, the spread of the luminous part of the opposite side
is caused by the light bending (gravitational lensing).
Such a spread can not be seen in the SR-calculation and the NR-calculation
since the light bending effect is not taken into consideration.
The orbit of the light is drastically bent around the NS
so that the opposite side of the column appears to be more widened
for the case of $\varphi_{\rm B}=0^\circ$ (panel a)
than for $\varphi_{\rm B}=180^\circ$ (panel b).
In addition, $\boldsymbol{v}_r\cdot\boldsymbol{n}$ tends to decrease
via the light bending, leading to the reduction of the radiation intensity on the screen.
The gravitational redshift also works to reduce the radiation
intensity for the GR-calculation.
Hence neither red nor yellow regions appear in the panels (a) and (b).

\subsection{Pulse Shape}
\label{sec:Light Curve}
\begin{figure}
 \centering
 \includegraphics[width=8cm]{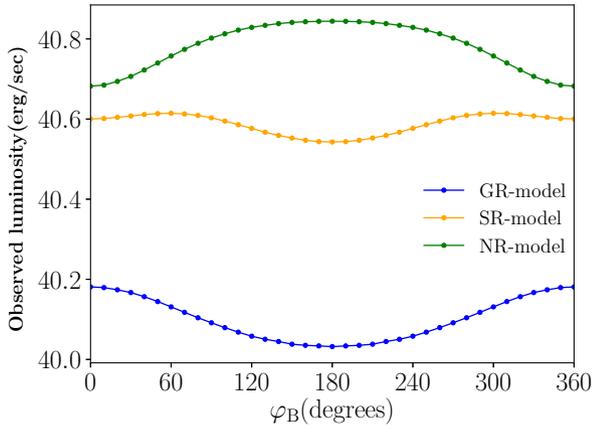}
 \caption{Pulse shape of GR-, SR-, and NR-calculations
   for the case of $\theta_\mathrm{B}=10^\circ$ and $\theta_\mathrm{obs}=60^\circ$.
   }
   \label{fig:LightCurve_log}
\end{figure}

The observed luminosity of the fiducial model
is plotted in figure \ref{fig:LightCurve_log} as a function of $\varphi_{\rm B}$.
As in figure \ref{fig:Imaging}, we employ
$\theta_\mathrm{B}=10^\circ$ and $\theta_\mathrm{obs}=60^\circ$.
This figure corresponds to the pulse shape
because $\varphi_{\rm B}$ periodically changes with time via the rotation of the NS.
The result of the GR-calculation, SR-calculation and NR-calculation are shown in blue, yellow,
and green circles, respectively.
As have mentioned in subsection \ref{sec:Imaging},
the radiation intensity tends to be reduced by the Doppler effect
since the redshift is proportional to $\gamma$.
Thus, the luminosity for the SR-calculation is smaller than that for the NR-calculation.
The gravitational redshift also works to reduce the luminosity
so that the luminosity for the GR-calculation is the smallest
although it still exceeds $10^{40}~{\rm erg~s}^{-1}$.

In this figure, we also find that
the observed luminosity depends on $\varphi_{\rm B}$ in all calculations.
However, the phases seem to be shifted by 180 degrees.
At $\varphi_{\rm B}=180^\circ$,
the luminosity is lowest in GR- and SR-calculations but highest in NR-calculation.
The reason for this is the Doppler effect.
The radiation intensity of the opposite side column becomes strong
due to the Doppler effect, especially when $\varphi_{\rm B}=0^\circ$.
Such enhancement is less effective as $\varphi_{\rm B}$ approaches $180^\circ$.
Therefore, the luminosity of the GR- and SR-calculations is minimized at $\varphi_{\rm B}=180^\circ$.
In the NR-calculation, the luminosity is only determined by the apparent area
of the luminous part of the accretion column.
Thus, the observed luminosity decreases
as $\varphi_{\rm B}$ approaches $0^\circ$ (or $360^\circ$).

In the GR-calculation, in addition to the Doppler effect,
the bright region in the opposite side of the accretion columns is widened
by the light bending effect, which leads to the enhancement of
the observed luminosity.
The light bending effect plays an important role
for the case of $\varphi_{\rm B} \sim 0^\circ$.
Thus, the luminosity for $\varphi_{\rm B}=0^\circ$ is maximum.
Note that,
if large $\theta_{\rm B}$ is employed ($\geq 70^\circ$),
the azimuthal angle of the magnetic axis, $\varphi_{\rm B}$,
where the observed luminosity becomes minimum,
is deviated from $180^\circ$.
This will be described in detail in subsection \ref{sec:PF}.

\subsection{Pulsed Fraction}
\label{sec:PF}
\begin{figure}
 \centering
 \includegraphics[width=8cm]{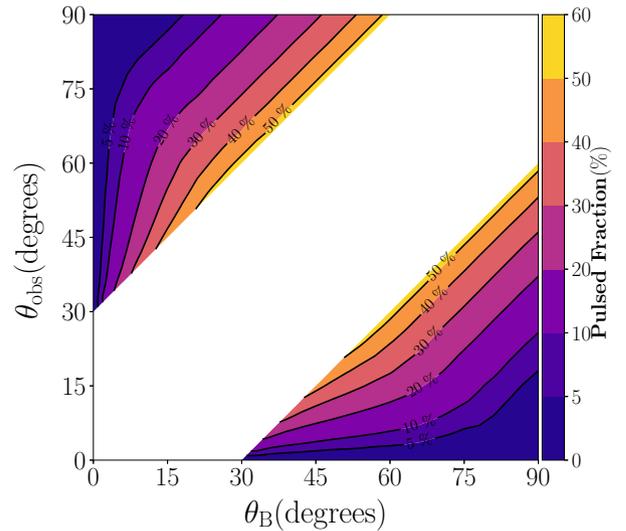}
 \caption{Pulsed fraction of the GR-calculation for various $\theta_{\rm B}$ and $\theta_{\rm obs}$.
 }
 \label{fig:PF_map1}
\end{figure}

Next, we show the PF, which is defined as
\begin{eqnarray}
 \mathrm{PF}=\frac{L_\mathrm{max}-L_\mathrm{min}}{L_\mathrm{max}+L_\mathrm{min}}.
\end{eqnarray}
Here, $ L_\mathrm{max}$ and $ L_\mathrm{min}$
are the maximum and minimum observed-luminosity in one period of the pulse, respectively.
In the case of $\theta_{\rm B}=10^\circ$ and $\theta_{\rm obs}=60^\circ$
(we have shown in subsection \ref{sec:Imaging} and \ref{sec:Light Curve}),
$L_{\rm max}=1.52 \times 10^{40}~{\rm erg~s}^{-1}$
and $L_{\rm min}=1.08 \times 10^{40}~{\rm erg~s}^{-1}$ in the GR-calculation,
so the PF is evaluated to be $17.0\%$.
We plot the PF with various $\theta_{\rm B}$ and $\theta_{\rm obs}$
on $\theta_{\rm B}$-$\theta_{\rm obs}$ plane (see figure \ref{fig:PF_map1}).

We find in this figure that the PF becomes less than 5\% at minimum
and exceeds 50 \% at maximum.
This figure also shows that
the PF tends to increase as
the observer's viewing angle, $\theta_{\rm obs}$, approaches
to $\theta_{\rm B} \pm \Theta_{\rm op}$ (the boundary of the white region),
 where $\Theta_{\rm op}$ is the half opening angle of columns (see section \ref{sec:model}).
This is because that
the relativistic effect effectively works
and the amplitude of the pulse is enhanced
as $\theta_{\rm obs}$ approaches to $\theta_{\rm B} \pm \Theta_{\rm op}$
(see subsection \ref{sec:Light Curve}).

The PF gradually decreases away from the white region,
and becomes very small
in the vicinity of the horizontal and vertical axis.
In the case of small $\theta_{\rm B}$,
the magnetic axis is almost aligned with the rotation axis.
Then, the observed luminosity does not change so much
(the amplitude of the resulting pulse becomes small).
Also, even if $\theta_{\rm B}$ is not small,
the amplitude of the pulse tends to be small
for the face-on observer (small $\theta_{\rm obs}$).
Thus, the PF is small at the region of
small $\theta_{\rm B}$ and small $\theta_{\rm obs}$.

Here we note that
the PF in the white region i.e., blank region) cannot be calculated in the present model.
In the blank region,
the offset angle between the magnetic axis and the observer's angle
becomes less than $\Theta_{\rm op}$
when $\varphi_{\rm B}$ is close to $0^\circ$.
That is,
the line of sight is inside the side boundary of
the accretion column, and we can not see the side wall of the accretion column.
This is a limitation of the present model,
in which we suppose that accretion column is the cone-shaped and infinitesimally-long.
In fact, it is thought that the accretion column bends
along the magnetic field lines, and therefore
the X-ray pulse might be observed even in the blank region.
The study of more realistic structure of the accretion column
is left as future work.
Global radiation magnetohydrodynamics simulations of
the accretion columns and accretion disks around the magnetized NSs
were recently performed
by \citet{Takahashi17}.
In addition, magnetohydrodynamics simulations showed the column accretion flows onto NSs
(\cite{Romanova09}; \yearcite{Romanova11}; \yearcite{Romanova12}; \cite{Parfrey}; \cite{Parfrey17}).

\begin{figure}
  \centering
  \includegraphics[width=8cm]{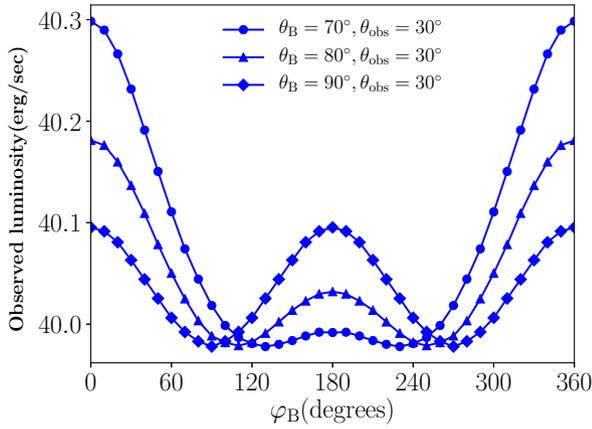}
  \caption{Pulse shape of the GR-calculation
    for the case of $\theta_{\rm B}=70^\circ$, $80^\circ$ and $90^\circ$.
    Here $\theta_{\rm obs}$ is set to be $30^\circ$.}
  \label{fig:LightCurve_obsgt60}
\end{figure}

We also note that a secondary peak ($\varphi_{\rm B}=180^\circ$)
appears during one period of the pulse
for the case that $\theta_{\rm B}$ is very large, $\gtsim 70^\circ$
(see figure \ref{fig:LightCurve_obsgt60}).
For the case of such larger $\theta_{\rm B}$,
the relativistic effect works to enhance the observed luminosity
when $\varphi_{\rm B}$ is close to null and $180^\circ$,
and the minimum luminosity appears between the primary peak
({$\varphi_{\rm B}=0^\circ, 360^\circ$)
and  the secondary peak ($\varphi_{\rm B}=180^\circ$).
Although the peak luminosity ($\varphi_{\rm B}=0^\circ$)
decreases with an increase of $\theta_{\rm B}$,
the minimum value of the luminosity is almost constant.
Therefore, the PF decreases with an increase of $\theta_{\rm B}$.
Especially, it is found that
the observed luminosity for $\varphi_{\rm B}=180^\circ$
is same as that for $\varphi_{\rm B}=0^\circ$
when $\theta_{\rm B}=90^\circ$.
In this case,
the period of the X-ray pulse becomes one half of
the rotation period of the NS.
The secondary peak in the pulse also appears
when observer is located near the equatorial plane
($\theta_{\rm obs} \gtsim 70^\circ$).
Our results propose that the angle of the magnetic axis
or the observer's viewing angle might
be restricted by the detailed observations of the pulse shape.

\subsection{Pulsed fraction of comparison models}
\label{sec:comparison models}
So far, we investigated the PF based on the simulation results of K16.
Here we show that our results are not so sensitive to the structure of the accretion column.
Figure \ref{fig:PF_map2} is the PF where we employ thinner accretion columns
of which the half opening angle, $\Theta_{\rm op}$, is $10^\circ$.
Here, the profiles of the temperature and the velocity of the side wall of the column
are the same as those of the fiducial model.
The intrinsic luminosity of this model is $2.1 \times 10^{40}~{\rm erg~s}^{-1}$.
Such thinner column would appear when the NS is strongly magnetized.
The accretion disk is truncated at a large distance because of the strong magnetic pressure.
The large magnetosphere is formed.
The matter falls onto the very vicinity of the magnetic pole,
forming to the thinner accretion column.
As shown in figure \ref{fig:PF_map2},
the PF is very small around the horizontal and vertical axis
and increases as approaching to the blank area.
This tendency is similar with the PF shown in figure \ref{fig:PF_map1},
although the blank region is narrow due to the small $\Theta_{\rm op}$.
Also, the maximum value of the PF does not change so much even if the half opening angle of the column is somewhat small or large.
When it is considered that both $\theta_{\rm B}$ and $\theta_{\rm obs}$ are  fixed, the PF is larger for $\Theta_{\rm op}=30^\circ$ (figure \ref{fig:PF_map1}) than $\Theta_{\rm op}=10^\circ$ (figure \ref{fig:PF_map2}).
For instance, for the case of $\theta_{\rm B}=15^\circ$ and $\theta_{\rm obs}=60^\circ$, the PFs are around 25 \% for $\Theta_{\rm op}=30^\circ $ and 15 \% for $\Theta_{\rm op}=10^\circ$.
Thus, the observed PF would change if $\Theta_{\rm op}$ varies for some reason (see section \ref{sec:discussion}).


\begin{figure}[t]
  \centering
  \includegraphics[width=8cm]{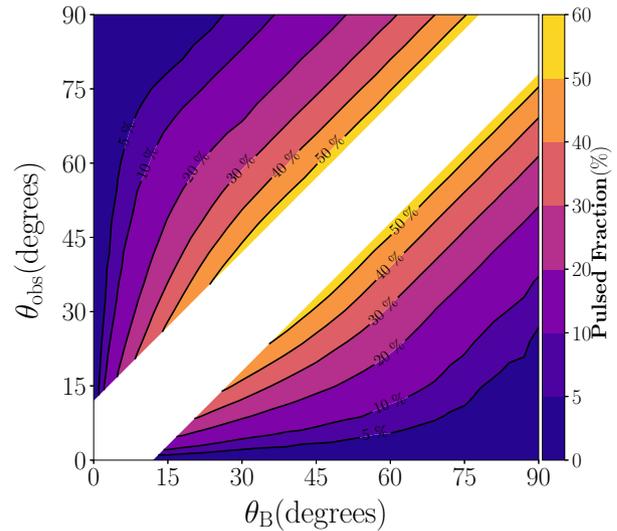}
  \caption{
    Same as figure \ref{fig:PF_map1} but
    we employ thinner accretion columns
    of which the opening angle is $10^\circ$.
  }
  \label{fig:PF_map2}
\end{figure}

Although the height of the shock surface is around 3 km in K16,
\citet{Basko76} indicated that
the shock surface moves in a direction away from the NS surface
with an increase of the mass accretion rate.
In order to study the influence of the height of the shock surface,
we set the height of the shock surface to be 6 km here.
Specifically,
without changing the absolute value of the temperature and the velocity,
the distance from the neutron star surface of the data points is doubled
(In other words,　$A^{\prime}(h)=A(0.5h)$, where $A^{\prime}(h)$ and $A(h)$
are physical quantity in this comparison model and in the fiducial model, respectively).
Then, the intrinsic luminosity is roughly doubled, $1.2 \times 10^{41}~{\rm erg~s}^{-1}$.
The resulting PF is plotted in figure \ref{fig:PF_map3}.
This figure is very similar with figure \ref{fig:PF_map1},
so that the PF is not sensitive to the position of the shock surface
as long as the shock occurs near the NS surface.
Although we employ simple comparison models in the present work,
the detailed structure of accretion columns
was investigated by \citet{Lyubarskii} and \citet{Mushtukov}.
The comparison of PFs between the present model and these latter detailed models
is indeed very important,
and we would like to consider this for our future work.

\begin{figure}[t]
  \centering
  \includegraphics[width=8cm]{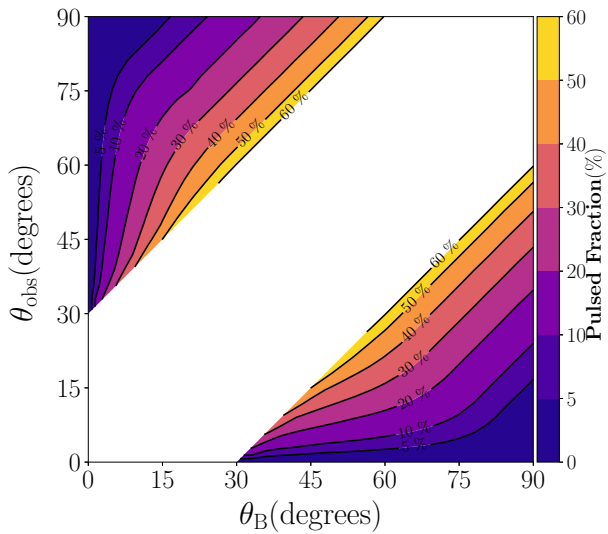}
  \caption{
    Same as figure \ref{fig:PF_map1} but
    the shock surface is set higher (6 km).
  }
  \label{fig:PF_map3}
\end{figure}

\section{Conclusions and Discussion}
\label{sec:discussion}
In the present study,
by taking account of the light bending effect, the doppler effect, and the gravitational redshift,
we calculate the luminosity of the super-critical column accretion flows onto the magnetized NSs
of which the structure is based on the numerical simulations by K16.
Due to the doppler effect and the light bending effect,
the radiation from the opposite side of the accretion column is enhanced.
Also the gravitational redshift decreases the observed luminosity.
The resulting luminosity highly exceeds the Eddington luminosity of the stellar-mass black holes.
If the magnetic axis of the NS is misaligned with the rotation axis,
the observed luminosity periodically changes.
The resulting PF tends to increase
as $\theta_{\rm obs}$ approaches to $\theta_{\rm B} \pm \Theta_{\rm op}$.
It exceeds $\sim 50$ \% for the case of
$\theta_{\rm obs} \sim \theta_{\rm B} \pm \Theta_{\rm op}$
in our calculations.
In contrast, it becomes less than 5 \%
in the case of $\theta_{\rm B}\lesssim 5^\circ$ or $\theta_{\rm obs}\lesssim 5^\circ$.

Our results are consistent with the observations of ULXPs with few exception mentioned in the next paragraph.
The PF of M82 X-2, NGC7793 P13, NGC5907 ULX, NGC1313 X-2, and M51 ULX7
was reported to be about
5-25 \% at 3-30 keV (\cite{Bachetti}), 8-40 \% at 1-15 keV (\cite{Furst}; \cite{Israelb}), 12-20 \%
at 2.5-7 keV (\cite{Israela}),
3-8 \% at 0.3-10 keV (\cite{Sathyaprakash}) and 5-40 \% at  0.3-9 keV (\cite{Rodriguez}),
respectively.
NGC300 ULX1 exhibits the PF of $\leq 50$ \% at the energy band of $\lesssim 1$ keV
\citep{Carpano}.
In the case of Swift J0243.6+6124,
three observations indicated the PF to be $\lesssim 40$ \%
\citep{Lian}.
Thus our results support the hypothesis
that the ULXPs are powered by the
the super-critical column accretion onto NSs.

However, some observations cannot be resolved in the present work.
The PF of $>50$ \% was detected at $\gtsim 1$ keV
in NGC300 ULX1 and two observations of Swift J0243.6+6124
showed the PF of $\gg 50$ \%.
In addition, it was reported that
the PF depends on the photon energy (see e.g., \cite{Bachetti, Wilson}).
The energy dependence of the PF and the high PF ($\gg 50$ \%) might be resolved
by the thermal and bulk Comptonization, which are not
taken into account in the present study.
Since the gas temperature as well as the infall velocity
are very high in the interior of the accretion column,
the emergent spectra might be deviated from the blackbody radiation
via the Comptonization.
The energy-dependent PF
is obtained by
solving the multi-frequency radiation transfer calculations including the Comptonization
\citep{Kawashima12, Kitaki, Narayan}.
Such study is left as an important future work.

Our results indicate that the detailed observations of the pulse shape might be useful for restricting the offset angle of the magnetic axis and the rotation axis as well as the observer’s viewing angle.
As we have shown in section \ref{sec:PF}, the secondary peak appears in a period of the pulse for $\theta_{\rm obs}>70^\circ$ or $\theta_{\rm B}>70^\circ$.
In contrast, in the situation other than that, the sinusoidal-shaped pulse appears.
This implies that the magnetic axis is not drastically misaligned with the rotation axis and that observer's viewing angle is relatively small in most ULXPs, 
since the sinusoidal pulse has been detected (see e.g., \cite{Bachetti}; \cite{Furst}; \cite{Israela}, \yearcite{Israelb}).
An exceptional object is Swift J0243.6+6124 of which the pulse has a secondary peak \citep{Wilson}. Thus, $\theta_{\rm obs}>70^\circ$ or $\theta_{\rm B}>70^\circ$ might be realized in this object.

Although the PFs are treated as constant in time in the present
study, the time variation of the PF and the intermittent signal has
been recently reported by \citet{Rodriguez} and \citet{Sathyaprakash}. The time variation of the PF might be
explained by the variation of the mass accretion rate. The increase
(decrease) of the mass accretion rate leads to the decrease (increase)
of the radius of the magnetosphere. Then the PF goes up (down) since
the half opening angle of the column becomes large (small). As have
shown in section \ref{sec:comparison models}, the PF tends to be large for the case with large
$\Theta_{\rm op}$. However, the PF might decrease (increase) due to the increase of
the mass accretion rate. The luminosity of the accretion disk, which
is located outside the magnetosphere, increases (decreases) with the
increase (decrease) of the mass accretion rate. The fraction of the
radiation from the columns in the total luminosity relatively decreases
(increases). Therefore, the PF decreases (increases) since the disk
does not exhibit the pulsed emission. This hypothesis is roughly
consistent with \citet{King17}. They have suggested that PF becomes
small if the magnetospheric radius is much smaller than the
spherization radius (see also \cite{Walton}).
Such a condition tends to be realized for the case that the mass accretion rate is large.
If the outflows are launched when the accretion rate increases and obscure the accretion column, 
the PFs might reduce or the X-ray pulse may disappear (\cite{Lian}).
We stress again that we need global radiation magnetohydrodynamics simulations of the accretion columns, accretion disks, and the outflows around the magnetized NS in order to reveal the PF by taking all effects mentioned above.
In addition, the reflection of light at the NS surface might reduce the PF
since the photons tend to be scattered towards the various directions.
The effect of the reflection was discussed by \citet{Lenzen} and \citet{West}
although we do not consider the reflection.
The study of the observed luminosity and the PF by taking account of the reflection
is also left as a future work.

Finally, we do not consider the quadrupole component of the magnetic field on the neutron star surface in the present paper.
Since the dipole component is dominant at the distant region, the disk matter would flow towards north/south poles.
However, the flow structure might change near the NS surface by the quadrupole component.
The investigation of the effect of the quadrupole fields is left as an important future work.

\bigskip
\begin{ack}
We thank an anonymous referee for fruitful comments. 
This work was supported by JSPS KAKENHI Grant Numbers JP17H01102A, JP18H04592, JP18K03710 (K.O.), JP18K13594, JP19H01906 (T.K.). 
This work was also supported in part by MEXT as a priority issue (Elucidation of the fundamental laws and evolution of the universe) 
to be tackled by using post-K Computer and JICFuS. 
\end{ack}

\end{document}